\documentclass[12pt]{article}
\usepackage{graphicx}
\usepackage{amsmath}
\usepackage{amssymb}
\usepackage{caption2}
\begin{document}
\bibliographystyle{prsty}
\begin{center}
{\large {\bf \sc{  decay constants of the  pseudoscalar charmonium and  bottomonium }}} \\[2mm]
Zhi-Gang Wang$^{1}$ \footnote{Corresponding author; E-mail,wangzgyiti@yahoo.com.cn.  }, Wei-Min Yang$^{2}$ and Shao-Long Wan$^{2} $    \\
$^{1}$ Department of Physics, North China Electric Power University, Baoding 071003, P. R. China \\
$^{2}$ Department of Modern Physics, University of Science and Technology of China, Hefei 230026, P. R. China \\
\end{center}

\begin{abstract}
In this article, we investigate the structures of the pseudoscalar
charmonium and  bottomonium in the framework of
 the coupled rainbow Schwinger-Dyson equation and ladder
 Bethe-Salpeter equation with the confining effective potential (infrared modified flat bottom potential).
 As the current masses are very large, the dressing or renormalization  for the $c$ and $b$ quarks
  are tender, however,
 mass poles in the timelike region are  absent.
The Euclidean time fourier transformed quark propagator has no
mass poles in the timelike region which
 naturally implements  confinement.
 The Bethe-Salpeter wavefunctions for those mesons
 have the same type (Gaussian type) momentum dependence
 and center around  zero momentum with spatial extension to about
$q^2=1GeV^2$ which happen to be the energy scale for Chiral
symmetry breaking, the strong interactions  in the infrared region
result in  bound states. The  decay constants for those
pseudoscalar heavy quarkonia are compatible with the values of
experimental extractions and theoretical calculations.
\end{abstract}


\section{Introduction}
Heavy quarkonium, bound state of the heavy quark and antiquark,
characterized by at least three widely separated energy scales:
the hard scale (the mass $m$ of the heavy quarks), the soft scale
(the relative momentum of the heavy quark--antiquark $|{\bf p}|$ )
and the ultrasoft scale ( the typical kinetic energy of the heavy
quark-antiquark $E$), plays a special role in probing the strong
interactions in both the perturbative and nonperturbative regions.
 By definition of the heavy quark, $m$
is large in comparison with the typical hadronic scale
$\Lambda_{QCD}$, the corresponding  processes  can be successfully
described in perturbative quantum chromodynamics (QCD)  due to the
asymptotic freedom. However, the lower scales   $|{\bf p}|$ and
$E$, which are responsible for the binding, can not be accessible
by perturbation theory.  The appearance of multiscales in the
dynamics of the heavy quarkonium makes its quantitative study
extremely difficult,  the properties of the  bound states and
their decays can provide  powerful   test for QCD in both the
perturbative and nonperturbative regions.

 The physicists propose many original approaches to deal with the long
distance properties of QCD, such as Chiral perturbation theory
\cite{Gasser85}, heavy quark effective theory \cite{Neubert94},
QCD sum rules \cite{Shifman79}, lattice QCD \cite{Gupta98},
perturbative QCD \cite{Brodsky80}, coupled Schwinger-Dyson
equation (SDE) and Bethe-Salpeter equation (BSE) method
\cite{Roberts94}, nonrelativistic QCD \cite{BBL95}, potential
nonrelativistic QCD \cite{Pineda04},  etc. All of those approaches
have both outstanding advantages and obvious shortcomings in one
or other ways.   The coupled rainbow SDE and ladder BSE have given
a lot of successful descriptions of the long distance properties
of the low energy QCD and the QCD vacuum (for example, Refs.
\cite{DHL,MarisRoberts97,MarisTandy99,Ivanov99} , for recent
reviews one can see Refs.\cite{Roberts00,Roberts03}). The SDE can
naturally  embody the dynamical symmetry breaking and confinement
which are two crucial features of QCD, although they correspond to
two very different energy scales \cite{Miransky93,Alkofer03}. On
the other hand, the BSE is a conventional approach in  dealing
with the two body relativistic bound state problems \cite{BS51}.
From the solutions of the BSE, we can obtain useful information
about the under-structure of the mesons and   obtain powerful
tests for the quark theory. However, the obviously drawback may be
the model dependent kernels for the gluon two-point Green's
function and the truncations for the coupled divergent SDE and BSE
series in one or the other ways\cite{WYW03}. Many analytical and
numerical calculations indicate that the coupled rainbow SDE and
ladder BSE with phenomenological potential models can give model
independent results and
 satisfactory values \cite{Roberts94,DHL,MarisRoberts97,MarisTandy99,Ivanov99,Roberts00,Roberts03}. The usually used
effective potential models are confining Dirac $\delta$ function
potential, Gaussian  distribution potential and flat bottom
potential (FBP)
\cite{Roberts00,Roberts03,Munczek83,Munczek91,Wangkl93}. The FBP
is a sum of Yukawa potentials, which not only  satisfies  chiral
invariance and fully relativistic covariance, but also suppresses
the singular point that the
 Yukawa potential has. It works well in
 understanding the dynamical chiral symmetry breaking, confinement and the QCD vacuum as well as
the meson  structures, such as electromagnetic form factors,
radius, decay constants \cite{WYW03,WangWan,Wan96}.

During  the past two years, the experiments have discovered a
number of new  states, for example, the  $\eta_{c}^{\prime}$ in
exclusive $B \to K K_{S} K^{-}\pi^{+}$ decays by Belle
\cite{Belle02} , the narrow $D_{sJ}$ states  by Babar, CLEO and
Belle \cite{BabarCLEOBelle}, evidence for the $\Theta^+(1540)$
with quantum numbers of $K^+n$  \cite{Theta03}, and  the $X(3872)$
through decay to $ \pi^+\pi^-J/\psi$ by Belle \cite{Belle03}. New
experimental results  call for interpretations, offer
opportunities to extend our knowledge about hadron spectrum and
challenge our understanding of the strong interaction;
  furthermore,  they  revitalize the study of
heavy quarkonia  and stimulate a lot of theoretical analysis
through the charmonia and bottomonia have been thoroughly
investigated.

The  decay constants of the pseudoscalar charmonium and
bottomonium ($\eta_c$ and $\eta_b$)
 mesons  play an important role in modern physics with the
assumption of current-meson duality. The precise knowledge of the
those values  $f_{\eta_c}$ and $f_{\eta_b}$ will provide great
improvements  in our understanding of various processes convolving
the $\eta_c$ and $\eta_b$ mesons, for example, the precess
$B\rightarrow \eta_c K$, where the mismatches between the
theoretical and experimental values are large \cite{WangLi}. The
 $\eta_c$ meson is already observed
experimentally,  the current experimental situation with the
 $\eta_b$ meson is rather uncertain,  yet the
discovery of the $\eta_b$ meson is one of the primary goals of the
CLEO-c research program \cite{Stock03}; furthermore, the $\eta_b$
meson   may be observed in Run II at the Tevatron through the
decay modes into charmed states $D^* D^{(*)}$\cite{Polosa04}. It
is interesting to combine those successful potential models within
the framework of coupled SDE and BSE to calculate the decay
constants of the pseudoscalar heavy quarkonia such as $\eta_c$ and
$\eta_b$ . For previous studies about the electroweak decays of
the pseudoscalar mesons with the SDE and BSE, one can consult
Refs.
\cite{Roberts94,DHL,MarisRoberts97,MarisTandy99,Ivanov99,Roberts00,Roberts03}.
In this article, we use an infrared modified flat-bottom potential
(IMFBP) which takes
 the advantages of
both the Gaussian distribution potential and the FBP to calculate
the decay constants of those  pseudoscalar heavy quarkonia.

The article is arranged as follows:  we introduce the IMFBP in
section II; in section III, IV and V, we solve the rainbow SDE and
ladder BSE, explore the analyticity of the quark propagators,
investigate the dynamical dressing  and confinement, finally
obtain the decay constants for those pseudoscalar  heavy
quarkonia; section VI is reserved for conclusion.

\section{Infrared modified Flat Bottom Potential }
The present techniques in QCD  calculation can not give
satisfactory large $r$ behavior for the gluon two-point Green's
function to implement the linear potential confinement  mechanism
, in practical calculation, the phenomenological effective
 potential models always do the work.
As in our previous work \cite{WYW03}, we use a gaussian
distribution function to represent the infrared behavior of the
gluon two-point Green's function,
\begin{eqnarray}
4\pi G_{1}(k^2)=3\pi^2 \frac{\varpi^2}{\Delta^2}e^{-\frac{k^2}{\Delta}},
\end{eqnarray}
which determines the quark-antiquark interaction through a
strength parameter $\varpi$ and a range  parameter $\Delta$.
 This form is inspired by the $\delta$ function potential
(in other words the infrared dominated potential) used in Refs.\cite{Munczek83,Munczek91}, which it approaches in the limit
$\Delta\rightarrow 0$. For the intermediate momentum, we take the FBP as the best approximation and neglect
the large momentum contributions from the perturbative QCD calculations as the coupling constant at high energy
is very small.
The FBP is a sum of Yukawa potentials which is an analogy to the
exchange of a series of particles and ghosts with different
masses (Euclidean Form),
\begin{equation}
G_{2}(k^{2})=\sum_{j=0}^{n}
 \frac{a_{j}}{k^{2}+(N+j \rho)^{2}}  ,
\end{equation}
where $N$ stands for the minimum value of the masses, $\rho$ is their mass
difference, and $a_{j}$ is their relative coupling constant.
 Due to the particular condition we take for the FBP,
there is no divergence in solving the SDE.
In its three dimensional form, the FBP takes the following form:
\begin{equation}
V(r)=-\sum_{j=0}^{n}a_{j}\frac{{\rm e}^{-(N+j \rho)r}}{r}  .
\end{equation}
In order to suppress the singular point at $r=0$, we take the
following conditions:
\begin{eqnarray}
V(0)=constant, \nonumber \\
\frac{dV(0)}{dr}=\frac{d^{2}V(0)}{dr^{2}}=\cdot \cdot
\cdot=\frac{d^{n}V(0)} {dr^{n}}=0    .
\end{eqnarray}
The  $a_{j}$ can be  determined by solve the equations  inferred
from the flat bottom condition Eq.(4). As in  previous literature
\cite{WYW03,Wangkl93,WangWan,Wan96}, $n$ is set to be 9. The
phenomenological effective
 potential (IMFBP) can be taken as
 \begin{equation}
 G(k^2)=G_1(k^2)+G_2(k^2).
 \end{equation}
\section{Schwinger-Dyson equation}
The SDE can provide a natural
  framework for investigating the nonperturbative properties  of the
  quark and gluon Green's functions. By studying the evolution
  behavior and analytic structure of the dressed quark propagators,
  we can obtain valuable information about the dynamical dressing phenomenon and confinement.
 In the following, we write down the rainbow SDE for the quark propagator,
\begin{equation}
S^{-1}(p)=i\gamma \cdot p + \hat{m}_{c,b}+ 4\pi \int \frac
{d^{4}k}{(2 \pi)^{4}} \gamma_{\mu}\frac{\lambda^a}{2}
S(k)\gamma_{\nu}\frac{\lambda^a}{2}G_{\mu \nu}(k-p),
\end{equation}
where
\begin{eqnarray}
S^{-1}(p)&=& i A(p^2)\gamma \cdot p+B(p^2)\equiv A(p^2)
[i\gamma \cdot p+m(p^2)], \\
G_{\mu \nu }(k)&=&(\delta_{\mu \nu}-\frac{k_{\mu}k_{\nu}}{k^2})G(k^2),
\end{eqnarray}
and $\hat{m}_{c,b}$ stands for the current quark mass that
explicitly breaks chiral symmetry.

The full SDE for the quark propagator is a divergent series of
coupled nonlinear integral equations for the propagators and
vertexes, we have to make truncations in one or other ways. The
rainbow  SDE has given a lot of successful descriptions of the QCD
vacuum and low energy hadron phenomena
\cite{Roberts94,Roberts00,Roberts03,Miransky93,Alkofer03}, in this
article, we take the rainbow SDE. If we go beyond the rainbow
approximation, the bare vertex $\gamma_\mu\frac{\lambda^a}{2}$ has
to be  substituted by the full quark-gluon vertex
$\Gamma_\mu^a(qqg)$, which satisfies the Slavnov-Tayler identity.
In the weak coupling limit, $g^2\rightarrow 0$, two Feynman
diagrams contribute to the vertex $\Gamma_\mu^a(qqg)$ at  one-loop
level due to the non-Abelian nature of QCD i.e. the
self-interaction of gluons \cite{Davydychev01}. If we neglect the
contributions from the three-gluon vertex $\Gamma_\mu^a(ggg)$ and
retain an Abelian version, the
 vertex $\Gamma_\mu^a(qqg)$ can be taken as
 $\frac{\lambda^a}{2}\Gamma_\mu(qqp)$, where the vertex $\Gamma_\mu(qqp)$ is the
 quark-photon vertex which satisfies the
 Ward-Takahashi identity. In practical calculation, we can
take the vertex $\Gamma_\mu(qqp)$ to be the Ball-Chiu  and
Curtis-Pennington vertex \cite{BallChiu,Curtis} so as to avoid
solving the  coupled SDE for the vertex $\Gamma_\mu(qqp)$.
However, the nonperturbative properties of QCD at the low energy
region suggest that the SDEs are strongly coupled nonlinear
integral equations, no theoretical work has ever proven that the
contributions from the vertex $\Gamma_\mu^a(ggg)$ can be safely
neglected due to the complex Dirac and tensor structures. The one
Feynman diagram contributions version of the vertex
$\Gamma_\mu^a(qqg)$ i.e. neglecting the contributions from the
vertex  $\Gamma_\mu^a(ggg)$ in dressing the vertex
$\Gamma_\mu^a(qqg)$ is inconsistent with the Slavnov-Tayler
identity \cite{Davydychev01}. If we take the assumption that the
contributions from the vertex $\Gamma_\mu^a(ggg)$ are not
different greatly from the
  vertex $\Gamma_\mu^a(qqg)$, we can multiply the contributions from
the vertex $\Gamma_\mu^a(qqg)$ by some parameters which
effectively embody the contributions from the vertex
$\Gamma_\mu^a(ggg)$ \cite{BhagwatST}.

In this article, we assume that a Wick rotation to Euclidean
variables is allowed, and perform a rotation analytically
continuing $p$ and $k$ into the Euclidean region. The Euclidean
rainbow SDE can be projected into two coupled integral equations
for $A(p^2)$ and $B(p^2)$. Alternatively, one can derive the SDE
from the
 Euclidean path-integral formulation of the theory, thus avoiding
 possible difficulties in performing the Wick
 rotation $\cite{Stainsby}$ . As far as only numerical results are concerned,
  the two procedures are equal. In fact, the analytical  structures of quark propagators have
 interesting information about confinement, we will make detailed discussion about
the  $c$ and $b$ quarks propagators  respectively in  section V.

\section{Bethe-Salpeter equation}
The BSE is a conventional approach in dealing with the two body
relativistic bound state problems \cite{BS51}. The precise
knowledge about the quark structures of the mesons will result in
better understanding of their properties. In the following, we
write down the ladder BSE for the pseudoscalar quarkonia,
\begin{eqnarray}
S^{-1}_{+}(q+\frac{P}{2})\chi(q,P)S^{-1}_{-}(q-\frac{P}{2})=\frac{16
\pi }{3} \int \frac{d^4 k}{(2\pi)^4}\gamma_\mu \chi(k,P)
\gamma_\nu G_{\mu \nu}(q-k),
\end{eqnarray}
where $S(q)$ is the quark propagator, $G_{\mu \nu}(k)$ is the
gluon propagator, $P_\mu$ is the four-momentum of the center of
mass of the pseudoscalar quarkonia, $q_\mu$ is the relative
four-momentum between the quark and antiquark, $\gamma_{\mu}$ is
the bare quark-gluon vertex,  and
 $\chi(q,P)$ is the Bethe-Salpeter wavefunction (BSW) of the bound state.

We can perform the Wick rotation analytically and continue  $q$
and $k$ into the Euclidean region \footnote{To avoid possible
difficulties in performing the Wick rotation, one can derive the
BSE from the Euclidean path-integral formulation of the theory. }.
 In the lowest order approximation, the BSW $\chi(q,P)$ can be
 written as
\begin{eqnarray}
\chi(q,P)=\gamma_5 \left[ iF_1^{0}(q,P)+\gamma \cdot P F_2^{0}(q,P)
+\gamma \cdot q q\cdot P F_3^{1}(q,P)+i[\gamma \cdot q,\gamma \cdot P  ] F_4^{0}(q,P) \right].
\end{eqnarray}
 The ladder BSE can be projected into the following four coupled integral equations,
\begin{eqnarray}
\sum_j H(i,j)F_j^{0,1}(q,P)&=&\sum_j \int_0^{\infty}k^3dk
\int_0^{\pi}\sin^2 \theta K(i,j) ,
\end{eqnarray}
the expressions of the $H(i,j)$ and $K(i,j)$ are cumbersome and neglected here.

Here we will give some explanations for the expressions of
$H(i,j)$ . The $H(i,j)$'s are functions of the quark's
Schwinger-Dyson functions (SDF) $A(q^2+ P^2/4+ q \cdot P)$ ,
$B(q^2+P^2/4+ q \cdot P)$, $A(q^2+P^2/4- q \cdot P)$ and
$B(q^2+P^2/4- q \cdot P)$. The relative four-momentum $q$ is a
quantity in the Euclidean spacetime  while the center of mass
four-momentum $P$ must be continued to the Minkowski spacetime
i.e. $P^2=-m^2_{\eta_c,\eta_b}$, this results in the $q \cdot P$
varying throughout a complex domain. It is inconvenient to solve
the SDE at the resulting complex values of the quark momentum,
especially for the heavy quarks. As the dressing effect is minor,
we can expand $A$ and $B$ in terms of Taylor series of  $q \cdot
P$, for example,
\begin{eqnarray}
A(q^2+P^2/4+ q \cdot P)&=&A(q^2+P^2/4)+A(q^2+P^2/4)' q \cdot
P+\cdots. \nonumber
 \end{eqnarray}
The other problem is that we can not solve the SDE in the timelike
region as the two point gluon Green's function can not be exactly
inferred from the $SU(3)$ color gauge theory even in the low
energy spacelike region. In practical calculations, we can
extrapolate the values of $A$ and $B$ from the spacelike region
smoothly to the timelike region with suitable  polynomial
functions. To avoid possible violation with confinement in sense
of the appearance of pole masses $q^2=-m^2(q^2)$ in the timelike
region, we must be care in
  choosing the polynomial functions \cite{Munczek91}. For the
  $\eta_c$  meson, the mass is about $3.0 GeV$, the extrapolation to the timelike region
  with  the quantity $-m^2_{\eta_c}/4$ can be performed easily,
   however, the large mass of the $\eta_b$ meson makes
  the extrapolation  into the deep
  timelike region troublesome. Although the $\eta_b$ meson has not been observed experimentally yet,
  the theoretical calculations indicate that its mass is  about $9.4 GeV$  \cite{Kniehl04}.   As the
  dressed quark propagators
  comprise the notation of constituent quarks by providing  a mass $m(q^2)=B(q^2)/A(q^2)$, which
  corresponding to  the dynamical symmetry breaking phenomena for the light quarks. We can
 simplify the calculation greatly and avoid the problems concerning the
 extrapolations
 in solving the BSE by take the following propagator
 for the $c$ and $b$ quarks,
 \begin{equation}
 S^{-1}(q^2)=i \gamma \cdot q +M_{c,b} \, \, ,
 \end{equation}
where the $M_{c,b}$ is the Euclidean constituent quark mass with
$M^2_{c,b}=m^2_{c,b}(q^2)=q^2$  obtained from the solution of the
SDE Eq.(6).

 Finally we write down the normalization condition for
the BSW,
\begin{eqnarray}
N_c \int \frac{d^4q}{(2\pi)^4} Tr \left\{ \bar{\chi}
\frac{\partial S^{-1}_{+}} {\partial P_{\mu}}\chi(q,P) S^{-1}_{-}
+\bar{\chi} S^{-1}_{+} \chi(q,P) \frac{\partial S^{-1}_{-}}
{\partial P_{\mu}} \right\}=2 P_{\mu},
\end{eqnarray}
where $\bar{\chi}=\gamma_4 \chi^+ \gamma_4$.

\section{Coupled rainbow SDE and ladder BSE and the decay constants}
In this  section, we explore the coupled equations of the rainbow
SDE and ladder BSE for the pseudoscalar heavy quarkonia
numerically, the final results for the SDFs and BSWs can be
plotted as functions of the square momentum $q^2$.

In order to demonstrate the confinement of quarks, we have to
study the analyticity of SDFs for the  $c$ and $b$ quarks, and
prove  that there no mass poles on the real timelike  $q^2$ axial.
 In the following, we  take the Fourier transform
 with respect to the Euclidean time T
 for the scalar part ($S_{s}$) of the quark propagator \cite{Roberts94,Roberts00,Maris95},
 \begin{eqnarray}
 S^{*}_{s}(T)  =  \int_{-\infty}^{+ \infty} \frac{dq_{4}}{2 \pi} e^{iq_{4}T}
 \frac{B(q^2)}{q^2A^2(q^2)+B^{2}(q^2)}|_{ \overrightarrow{q}=0},
 \end{eqnarray}
where the 3-vector part of $q$ is set to zero.
 If  S(q) has a mass pole at $q^2=-m^2(q^2)$ in the real timelike region, the Fourier transformed
  $S^{*}_{s}(T)$ would fall off as $e^{-mT}$ for large T or
  $\log{S^{*}_{s}}=-mT$.

In our numerical calculations, for small $T$, the values of
$S^{*}_{s}$ are positive  and  decrease rapidly to zero and beyond
with the increase  of $T$, which are compatible with the result
(curve tendency with respect to $T$) from  lattice simulations
\cite{Bhagwat03} ; for large $T$, the values of $S^{*}_{s}$ are
negative, except occasionally a very small fraction of positive
values.  The negative values for $S^{*}_{s}$ indicate  an explicit
violation of the axiom of reflection positivity \cite{Jaffee},
 in other words, the quarks are not physical observable i.e. confinement.

For the $c$ and $b$ quarks, the current masses are very large, the
dressing or renormalization is
 tender and the curves are not steep which in contrast to the dynamical chiral
symmetry breaking phenomenon for the light quarks,
${m_c(0)}/{\hat{m}_c}\simeq 1.5$ and ${m_b(0)}/{\hat{m}_b}\simeq
1.1$, however, mass poles in the timelike region are absent. At
zero momentum, $m_c(0)= 1937 MeV $ and $m_b(0)=5105 MeV $, while
the Euclidean constituent quark masses $M_c= 1908 MeV $ and
$M_b=5096 MeV $, which defined by $M^2=m^2(q^2)=q^2$,  are
compatible with the constituent quark masses in the literature.
 From the plotted BSWs (see Fig.1 as an example),
we can see that the BSWs for  pseudoscalar mesons have the same type (Gaussian type) momentum dependence while
   the quantitative values are different from each other. Just like the lighter $\bar{q}q$ and $\bar{q}Q$
   pseudoscalar  mesons \cite{WYW03},
   the gaussian type BSWs
center around  zero momentum with spatial extension to about
$q^2=1GeV^2$ which happen to be the energy scale for Chiral
symmetry breaking, the strong interactions in the infrared region
result  in  bound states. Finally we obtain the values for the
decay constants of those pseudoscalar mesons which are defined by
\begin{eqnarray}
i f_{\pi} P_\mu &=& \langle0|\bar{q}\gamma_\mu \gamma_5 q |\pi(P)\rangle, \nonumber \\
&=& N_c \int Tr \left[\gamma_\mu \gamma_5\chi(k,P)\right]
\frac{d^4 k}{(2\pi)^4} ,
\end{eqnarray}
here we use $\pi$ to represent the pseudoscalar
mesons\footnote{Here we write down  the $N_c$ explicitly according
to the normalization condition Eq.(13). },
\begin{eqnarray}
f_{\eta_c}=349 MeV;  \, \, \, f_{\eta_b}=287 MeV,
\end{eqnarray}
which are compatible with the results from the experimental
extractions and theoretical calculations, $f_{\eta_c}=335\pm 75
MeV (Exp)$  \cite{etacExp01}; $f_{\eta_c}\approx 400 MeV (Exp)$
 \cite{PDG96};
 $f_{\eta_c}=420\pm52 MeV, f_{\eta_b}=705\pm 27 MeV (Theor)$
 \cite{Hwang97};
$f_{\eta_c}=292 \pm 25  MeV (Theor)$  \cite{Cvetic04}; $f_{\eta_c}
\approx 350 MeV (Theor)$ \cite{Novikov78}; $f_{\eta_c} = 300\pm50
MeV (Theor)$ \cite{Deshpande94}. In calculation, the values of
$\hat{m}_c$ and $\hat{m}_b $ are taken as the current quark
masses,  $\hat{m}_c=1250MeV$ and $\hat{m}_b=4700 MeV$; the input
parameters for the FBP are $N=1.0 \Lambda $, $V(0)=-11.0 \Lambda$,
 $\rho=5.0\Lambda$ and $\Lambda=200 MeV$, which are determined in  study of the  $\bar{q}q$
 and $\bar{q}Q$ pseudoscalar  mesons \cite{WYW03}. In this article,  the Euclidean
 constituent quark masses for  the $c$ and $b$ quarks are taken in solving the BSE as the dressing is tender. We
   borrow some idea  from the fact that
 the simple phenomenological model of Cornell potential (Coulomb potential plus
 linear potential) with constituent quark masses can give satisfactory mass
 spectrum for the heavy quarkonia\footnote{For an excellent review of the potential models,
 one can consult Ref.\cite{Quigg79}. } and  take larger values for
 the strength parameter $\varpi$ and  range parameter $\Delta$, i.e. $\varpi=2.2 GeV$ and  $\Delta=2.9
 GeV^2$,  in the infrared region comparing with
 the corresponding ones used in Ref.\cite{WYW03}.
  Furthermore  the masses of the
pseudoscalar mesons are taken as input parameters. If we take the
Euclidean constituent quark masses $M_c=m_c(0)$ and $M_b=m_b(0)$,
the decay constants for the $\eta_c$ and $\eta_b$ mesons change
slightly, $f_{\eta_c}=357 MeV$ and $f_{\eta_b}=289 MeV$.

\section{Conclusion }
In this article, we investigate the under-structures of the
pseudoscalar heavy quarkonia  $\eta_c$ and $\eta_b$ in the
framework of the coupled rainbow SDE and ladder BSE with the
confining effective potential (IMFBP). After we solve the coupled
rainbow SDE and ladder BSE numerically, we obtain the SDFs and
BSWs for the pseudoscalar heavy quarkonia $\eta_c$ and $\eta_b$.
As the current masses of the $c$ and $b$ quarks are very large,
the dressing or renormalization for the  SDFs  is  tender and the
curves are not steep which in contrast to the  explicitly
dynamical chiral symmetry breaking phenomenon for the light
quarks, however, mass poles in the timelike region are absent. We
can
 simplify the calculation greatly and avoid  the problems concerning the
 extrapolations
  in solving the BSE by making the
 substitution $B(q^2)\rightarrow M$ and $A(q^2)\rightarrow
 1$. The BSWs for the
pseudoscalar heavy quarkonia have the same type (Gaussian type)
momentum dependence while
   the quantitative values are different from each other. The gaussian type BSWs
    center around  zero momentum with spatial extension to about
$q^2=1GeV^2$ which happen to be the energy scale for Chiral
symmetry breaking, the strong interactions in the infrared region
result  in  bound states. Our numerical results for the values of
the decay constants of the pseudoscalar heavy quarkonia are
compatible with the corresponding ones obtained from the
experimental extractions and theoretical calculations. Once the
satisfactory SDFs and BSWs for   the pseudoscalar heavy quarkonia
are known, we can use them to investigate a lot of important
quantities involving the B, $\eta_c$ and $\eta_b$ mesons.

\section*{Acknowledgment}
This  work is supported by National Natural Science Foundation,
Grant Number 10405009,  and Key Program Foundation of NCEPU. The
authors are indebted to Dr. J.He (IHEP) for numerous help, without
him, the work would not be finished. One of the author (Z.G.Wang)
would like to thank
 Dr. Gogohia for helpful discussion.

\end{document}